\documentclass[smallcondensed]{svjour3}     
\smartqed  

\usepackage{graphicx}
\usepackage{lineno,hyperref}
\usepackage{multirow}
\usepackage{graphicx}
\usepackage{epstopdf}
\usepackage[cmex10]{amsmath}
\usepackage{amssymb}

\usepackage{multirow}
 \usepackage{algpseudocode}
\usepackage{algorithm}
\usepackage{amsthm}

\begin{document}

\title{Performance of Uplink NOMA with User Mobility Under Short Packet Transmission
\thanks{This work has been supported by the Personal Agreement 2014-2020, Greece; funded project for new researchers - phase II (code 5050174).}
}
\subtitle{}


\author{Nikolaos~I.~Miridakis \and Emmanouel~T.~Michailidis  \and Angelos~Michalas  \and Emmanouel~Skondras  \and Dimitrios~J.~Vergados  \and Dimitrios~D.~Vergados
}


\institute{N.~I.~Miridakis \at
              Department of Informatics and Computer Engineering, University of West Attica, Greece. \\
                            \email{nikozm@uniwa.gr}           
           \and
           E.~T.~Michailidis \at
              Department of Electrical and Electronics Engineering, University of West Attica, Greece.
							\and
           Angelos~Michalas \at
              Department of Electrical and Computer Engineering, University of Western Macedonia, Greece.
							\and
           E. Skondras \at
              Department of Informatics, University of Piraeus, Greece.
							\and
           Dimitrios~J.~Vergados \at
              Department of Informatics, University of Western Macedonia, Greece.
							\and
           Dimitrios~D.~Vergados \at
              Department of Informatics, University of Piraeus, Greece.
}

\date{Received: date / Accepted: date}

\maketitle

\begin{abstract}
The scenario of an uplink two-user non-orthogonal multiple access (NOMA) communication system is analytically studied when it operates in the short packet transmission regime. The considered users support mobility and each is equipped with a single antenna, while they directly communicate with a multi-antenna base station. Power-domain NOMA is adopted for the signal transmission as well as the successive interference cancellation approach is performed at the receiver for decoding. The packet error rate (PER) is obtained in simple closed formulae under independent Rayleigh faded channels and for arbitrary user mobility profiles. The practical time variation and correlation of the channels is also considered. Moreover, useful engineering insights are manifested in short transmission time intervals, which define a suitable setup for the forthcoming ultra-reliable and low latency communication systems. Finally, it turns out that the optimal NOMA power allocation can be computed in a straightforward cost-effective basis, capitalizing on the derived PER expressions.
\end{abstract}

\section{Introduction}
Non-orthogonal multiple access (NOMA) is considered as one of the most prominent solutions for $5$G systems due to its increased spectral efficiency over conventional orthogonal multiple access. To avoid jointly using time, frequency and/or code resources, power-domain NOMA has gained considerable attention via superposing multiple users' signal in the power domain. The received signals create mutual interference. Yet, the detector can employ successive interference cancellation (SIC), i.e., it decodes the strongest signal, and then subtracts it from the remaining signal, such that the next detection stage is interference-free. Doing so, NOMA can increase the system spectral efficiency and multiuser diversity as well as reduce latency. Nowadays, various Internet-of-things (IoT) applications utilize short packet transmission since connectivity and low-latency rather than high throughput is of prime importance. More so, ultra-reliable and low-latency communication (URLLC), which is at the forefront of wireless communications, defines another quite interesting and exciting paradigm that utilizes short packet transmission.

Due to the complementary benefits of NOMA and short packet transmission, their joint investigation has attracted a lot of research interest lately \cite{Amjad18}-\cite{HuangYang19}. In particular, coded NOMA was studied in \cite{Amjad18}-\cite{Ghanami21} when all the considered nodes are equipped with single antennas. In \cite{Tran21}-\cite{HuangYang19}, the case when a multi-antenna receiver is employed was analytically studied. Nevertheless, all the aforementioned works have not considered user mobility, which is present in various modern practical networking setups (e.g., vehicles and moving devices). To our knowledge, the scenario of NOMA short packet transmission systems with user mobility has not been investigated to date. Capitalizing on the mentioned observations, we analytically study a two-user uplink NOMA setup operating in the short packet transmission regime. The users support mobility and the time variation as well as correlation of the channel is modeled by the second-order statistic of level crossing rate (LCR) with respect to either the signal-to-interference-plus-noise ratio (SINR) or signal-to-noise ratio (SNR). Single-antenna mobile users and a multi-antenna base station are adopted, while the received signals undergo independent Rayleigh channel fading conditions. Unlike most previous research works, we capitalize on the fact that SINR (and SNR) is a stationary stochastic process and, doing so, a two-state Markov model is used to analyze the packet error rate (PER) for finite (short) packet transmissions. In addition, we show that the optimal NOMA power allocation can be numerically computed in a straightforward and cost-effective manner based on the derived PER expressions. 

\section{System Model}
Consider an uplink wireless communication setup, where two single-antenna users transmit to a base station equipped with $N$ antenna elements. Power-domain NOMA is used for the signal transmission so as to achieve spectral efficiency, while both signals undergo independent Rayleigh channel fading conditions. Let $h_{i}$ and $s_{i}$ denote the channel fading and transmitted symbols of the $i^{\rm th}$ user, respectively, with $i\in \{1,2\}$. The time autocorrelation of each channel path of the $i^{\rm th}$ user is given by $J_{0}(2\pi f_{i} \tau)$, where $J_{0}(\cdot)$ denotes the zeroth-order Bessel function, $\tau$ is the time difference between two correlated samples and $f_{i}\triangleq v_{i}/\lambda$ stands for the maximum Doppler frequency associated with this channel path. Also, $v_{i}$ and $\lambda$ represent the relative mobile speed of the $i^{\rm th}$ user and carrier wavelength, correspondingly. 

The received signal reads as
\begin{align}
\mathbf{r}=p \sum^{2}_{i=1}\alpha_{i} \mathbf{h}_{i} s_{i}+\mathbf{w},
\label{receivedSignal}
\end{align}
where $\mathbf{r}\in \mathbb{C}^{N\times 1}$ is the received signal vector, $p$ is the transmit power and $\mathbf{w}\in \mathbb{C}^{N\times 1}$ stands for the additive white Gaussian noise vector with its elements having zero-mean and variance $N_{0}$. Also, $\alpha_{i}$ defines the NOMA power allocator of the $i^{\rm th}$ user, such that $0<\alpha_{i}<1$ and $\alpha_{1}\triangleq 1-\alpha_{2}$. Further, $\mathbf{h}_{i}\in \mathbb{C}^{N\times 1}$ represents the received channel fading vector with complex Gaussian entries having zero-mean and unit-variance. Upon the signal reception, SIC is used for detection. 

Without loss of generality, we assume that $s_{1}$ is firstly detected by applying the weight vector $\mathbf{v}_{1}\triangleq \mathbf{h}_{1}/\|\mathbf{h}_{1}\|$ to the received signal; i.e., $\mathbf{v}^{\mathcal{H}}_{1} \mathbf{r}$, where the superscript $\mathcal{H}$ denotes the Hermitian transpose operator and $\|\cdot\|$ stands for the Euclidean (vector) norm. Afterwards, in the case when $s_{1}$ is correctly decoded and removed, the same procedure follows for $s_{2}$ to the remaining received signal via the weight vector $\mathbf{v}_{2}\triangleq \mathbf{h}_{2}/\|\mathbf{h}_{2}\|$, i.e., $\mathbf{v}^{\mathcal{H}}_{2} \mathbf{r}'$, where $\mathbf{r}'=\alpha_{2} \mathbf{h}_{2} s_{2}+\mathbf{w}$ is the remaining received signal after the contributing part of $s_{1}$ is stripped off. Thereby, assuming a unit-power signal $s_{i}$ and perfect channel state information at the receiver, the SINR and SNR at the $1^{\rm st}$ and $2^{\rm nd}$ SIC detection stage is given, respectively, as
\begin{align}
\gamma_{1}\triangleq \frac{p \alpha_{1} \|\mathbf{h}_{1}\|^{2}}{p \alpha_{2} |\mathbf{v}^{\mathcal{H}}_{1} \mathbf{h}_{2}|^{2}+N_{0}},
\label{g1}
\end{align}    
and
\begin{align}
\gamma_{2}\triangleq \frac{p \alpha_{2}}{N_{0}} \|\mathbf{h}_{2}\|^{2},
\label{g2}
\end{align} 
where $|\cdot|$, present in the denominator of \eqref{g1}, denotes the absolute (scalar) value operator.

\section{Performance Metrics}
We commence by analyzing PER of the considered uplink NOMA short packet transmission setup. First, we assume that a packet error occurs only when $\gamma_{i}<\gamma_{\rm th}$ for any time instance during an entire packet duration with $\gamma_{\rm th}$ denoting a certain data rate threshold (in bps/Hz). This process is modeled by using the two-state Markov model analyzed in \cite[\S III.B]{Fukawa12}. Thereupon, PER at the $1^{\rm st}$ detection stage becomes \cite[Eq. (58)]{Fukawa12}
\begin{align}
\overline{P^{(1)}_{s}}=1-\exp\left(-\frac{T_{\rm p}{\rm LCR}_{1}(\gamma_{\rm th})}{1-F_{\gamma_{1}}(\gamma_{\rm th})}\right)\left(1-F_{\gamma_{1}}(\gamma_{\rm th})\right),
\label{PER1}
\end{align}
where $T_{\rm p}$, ${\rm LCR}_{i}(\cdot)$ and $F_{\gamma_{i}}(\cdot)$ denote the packet transmission time interval, LCR and cumulative distribution function (CDF) of $\gamma_{i}$, respectively. Notably, the defined PER considers the time variations and correlations of the channel, captured by the second-order LCR statistic. Moreover, according to the adopted SIC-enabled reception approach, conditioned on the successful decoding of the $1^{\rm st}$ stage, the $2^{\rm nd}$ stage is being performed. Thus, the instantaneous packet error at the $2^{\rm nd}$ stage can be modeled by $P^{(1)}_{s}+P^{(2)}_{s,C}(1-P^{(1)}_{s})$, where $P^{(1)}_{s}$ and $P^{(2)}_{s,C}$ are the instantaneous error probability at the $1^{\rm st}$ detection stage and conditional error probability at the $2^{\rm nd}$ stage given that the $1^{\rm st}$ stage was error-free. Note that the conditional (average) PER of the $2^{\rm nd}$ stage, i.e., $\overline{P^{(2)}_{s,C}}$ is computed as per \eqref{PER1}, by simply substituting $F_{\gamma_{1}}$ and ${\rm LCR}_{1}$ with $F_{\gamma_{2}}$ and ${\rm LCR}_{2}$, correspondingly. Thereby, the following union bound on the unconditional PER at the $2^{\rm nd}$ stage yields as
\begin{align}
\overline{P^{(2)}_{s}}\leq \sum^{2}_{i=1} 1-\exp\left(-\frac{T_{\rm p} {\rm LCR}_{i}(\gamma_{\rm th})}{1-F_{\gamma_{i}}(\gamma_{\rm th})}\right)\left(1-F_{\gamma_{i}}(\gamma_{\rm th})\right)\triangleq \overline{P^{(2)}_{s,B}}.
\label{PER2}
\end{align}  
Typically, error rates remain considerably low in ultra-reliable applications, i.e., $\propto \{10^{-6},10^{-4}\}$; hence, the latter union bound is quite sharp, yielding $\overline{P^{(2)}_{s}}\approx \overline{P^{(2)}_{s,B}}$. 

According to \eqref{g1} and without delving into details, it is straightforward to show that the corresponding CDF of SINR is given by
\begin{align}
\nonumber
F_{\gamma_{1}}(\gamma_{\rm th})&=\int^{\infty}_{0}F_{\alpha_{1} \|\mathbf{h}_{1}\|^{2}}\left(y \gamma_{\rm th}+\frac{\gamma_{\rm th} N_{0}}{p}\right) f_{\alpha_{2} |\mathbf{v}^{\mathcal{H}}_{1} \mathbf{h}_{2}|^{2}}(y) dy\\
\nonumber
&=1-\sum^{N-1}_{k=0}\frac{\gamma^{k}_{\rm th} \exp\left(\frac{N_{0}}{p \alpha_{2}}\right)}{k! \alpha_{2} \left(\frac{\gamma_{\rm th}}{\alpha_{1}}+\frac{1}{\alpha_{2}}\right)^{k+1}}\Gamma\left(k+1,\frac{\alpha_{1}+\alpha_{2} \gamma_{\rm th}}{\alpha_{1} \alpha_{2} p/N_{0}}\right)\\
&=1-\sum^{N-1}_{k=0}\sum^{k}_{l=0}\frac{\gamma^{k}_{\rm th} \exp\left(-\frac{\gamma_{\rm th} N_{0}}{p \alpha_{1}}\right)}{l! \left(p/N_{0}\right)^{l}\alpha_{2} \left(\frac{\gamma_{\rm th}}{\alpha_{1}}+\frac{1}{\alpha_{2}}\right)^{k-l+1}},
\label{F1}
\end{align}  
where $\Gamma(\cdot,\cdot)$ is the upper incomplete Gamma function. Also, $F_{\alpha_{1} \|\mathbf{h}_{1}\|^{2}}(\cdot)$ and $f_{\alpha_{2} |\mathbf{v}^{\mathcal{H}}_{1} \mathbf{h}_{2}|^{2}}(\cdot)$ denote the Erlang CDF and exponential probability density function (PDF), respectively. In a similar basis, according to \eqref{g2}, it holds that 
\begin{align}
\nonumber
F_{\gamma_{2}}(\gamma_{\rm th})&=1-\frac{\Gamma\left(N,\frac{N_{0} \gamma_{\rm th}}{p \alpha_{2}}\right)}{\Gamma(N)}\\
&=1-\exp\left(-\frac{N_{0} \gamma_{\rm th}}{p \alpha_{2}}\right)\sum^{N-1}_{k=0}\frac{\left(\frac{N_{0} \gamma_{\rm th}}{p \alpha_{2}}\right)^{k}}{k!}.
\label{F2}
\end{align} 

On another front, LCR is an important second-order statistic that showcases the rate of fading occurrence within a certain time interval. According to the structure of $\gamma_{1}$ in \eqref{g1}, its corresponding LCR is expressed as \cite[Eq. (21)]{AliTorlak17}
\begin{align}
\nonumber
&{\rm LCR}_{1}(\gamma_{\rm th})=\\
\nonumber
&\frac{\sqrt{2 \pi} f_{1} \left(\frac{\gamma_{\rm th} N_{0}}{p \alpha_{1}}\right)^{N-\frac{1}{2}} \exp\left(-\frac{\gamma_{\rm th} N_{0}}{p \alpha_{1}}\right)}{\left(1+\frac{\gamma_{\rm th} \alpha_{2} f^{2}_{2}}{\alpha_{1} f^{2}_{1}}\right)^{N-1} \left(1+\frac{\gamma_{\rm th} \alpha_{2}}{\alpha_{1}}\right)\exp\left(\frac{\alpha_{2}p (\alpha_{1} f^{2}_{1}+\alpha_{2} f^{2}_{2} \gamma_{\rm th})}{f^{2}_{1} N_{0}(\alpha_{1}+\alpha_{2} \gamma_{\rm th})}\right)}\\
&\times \sum^{N-1}_{l=0}\frac{\left(\frac{\gamma_{\rm th} \alpha_{2} f^{2}_{2}}{\alpha_{1} f^{2}_{1}}\right)^{N-l-1} \Gamma\left(l+\frac{3}{2},\frac{f^{2}_{1} N_{0}(\alpha_{1}+\alpha_{2} \gamma_{\rm th})}{\alpha_{2}p (\alpha_{1} f^{2}_{1}+\alpha_{2} f^{2}_{2} \gamma_{\rm th})}\right)}{l! (N-l-1)!\left(\frac{f^{2}_{1} N_{0}(\alpha_{1}+\alpha_{2} \gamma_{\rm th})}{\alpha_{2}p (\alpha_{1} f^{2}_{1}+\alpha_{2} f^{2}_{2} \gamma_{\rm th})}\right)^{l+\frac{1}{2}}}.
\label{LCR1}
\end{align} 
Finally, the LCR of $\gamma_{2}$ is presented as \cite[Eq. (17)]{Beaulieu03}
\begin{align}
{\rm LCR}_{2}(\gamma_{\rm th})=\frac{\sqrt{2 \pi} f_{2} \left(\frac{\gamma_{\rm th} N_{0}}{p \alpha_{2}}\right)^{N-\frac{1}{2}}}{\Gamma(N) \exp\left(\frac{\gamma_{\rm th} N_{0}}{p \alpha_{2}}\right)}.
\label{LCR2}
\end{align} 

\section{Engineering Insights}
In the short packet transmission regime, recall that $T_{\rm p}$ remains quite low. In addition, when ultra-high reliability is required in the considered NOMA setup, it is obvious that a high SNR is a requisite. Capitalizing on the said observations and applying the first-order McLaurin series to the exponential function within \eqref{PER1} (i.e., $\exp(-z)\rightarrow 1-z$ as $z\rightarrow 0^{+}$), we arrive at  
\begin{align}
\overline{P^{(1)}_{s}}\approx F_{\gamma_{1}}(\gamma_{\rm th})+T_{\rm p}{\rm LCR}_{1}(\gamma_{\rm th}).
\label{PER1asy}
\end{align}
Likewise, PER at the $2^{\rm nd}$ stage approaches
\begin{align}
\overline{P^{(2)}_{s}}\approx F_{\gamma_{1}}(\gamma_{\rm th})+F_{\gamma_{2}}(\gamma_{\rm th})+T_{\rm p}\left[{\rm LCR}_{1}(\gamma_{\rm th})+{\rm LCR}_{2}(\gamma_{\rm th})\right].
\label{PER2asy}
\end{align}
It is noteworthy that the PER is lower bounded by the corresponding outage probabilities at each detection stage (say, $F_{\gamma_{i}}(\cdot)$), whereon an `\emph{extra penalty}' is added so as to reach the total PER, which is reflected by the corresponding LCR performance of $\gamma_{i}$.

The optimal power allocation, defined by $\alpha^{\star}$, defines a key performance indicator for the considered NOMA setup. Specifically, the optimization problem can be designed such that PER at the $2^{\rm nd}$ stage should be minimized under the constraint of PER at the $1^{\rm st}$ stage not exceeding a predetermined threshold value $\epsilon$. It can be formulated as
\begin{align}
\mathcal{P}_{1}: &\quad \underset{\alpha^{\star}}{{\rm arg\:min}} \:\overline{P^{(2)}_{s}}\\
\nonumber
& \textrm{subject to:  } \overline{P^{(1)}_{s}}\leq \epsilon,\quad 0<\alpha_{1}<1.
\end{align}
Unfortunately, $\mathcal{P}_{1}$ is a non-convex optimization problem since both the objective and constraint function (regarding PER) are non-convex functions. However, $\alpha^{\star}$ can be directly obtained by a numerical search over the real line in the range $\{0,1\}$ for arbitrary antenna arrays $N$ and user mobility profiles. 

\section{Numerical Results}
The derived analytical results are verified via numerical validation where they are cross-compared with corresponding Monte-Carlo simulations. The Rayleigh faded channels are generated by using the sum-of-sinusoids method \cite{Patzold09}, which is a modified Jakes model. In what follows, for ease of presentation and without loss of generality, we assume that $f_{1}=f_{2}\triangleq f$; namely, the two users have an identical mobility profile (i.e., the same relative speed). The packet duration $T_{\rm p}$ is set to be $1$ millisecond, appropriate for low-latency short packet transmission.

\begin{figure}[h]
\centering{\includegraphics[trim=0.5cm -0.0cm -0.0cm 0.0cm, clip=true,totalheight=0.45\textheight]{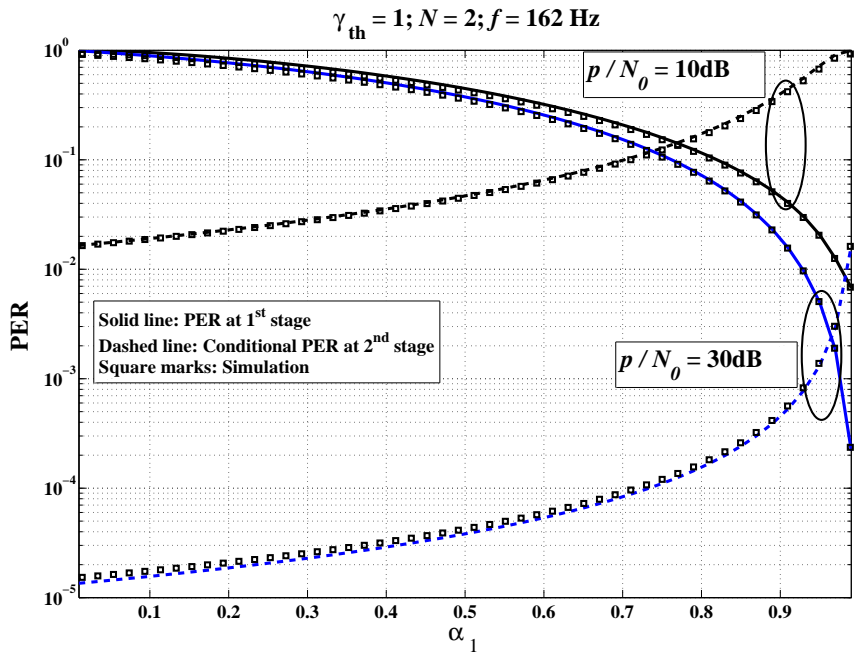}}
\caption{PER vs. various power allocation factors $\alpha_{1}$.}
\label{fig1}
\end{figure}

In Fig.~\ref{fig1}, the maximum Doppler frequency is set to be $f=162$ Hz, which corresponds to a vehicular speed of $50$ km/hr regarding a cellular system with carrier frequency $3.5$ GHz. We model a base station equipped with $2$ antenna elements operating in a high SNR region. The PER of the $1^{\rm st}$ detection stage is depicted as well as the conditional PER at the $2^{\rm nd}$ stage (given that the $1^{\rm st}$ stage is correctly decoded), which can be directly computed by setting $i=2$ in \eqref{PER2}. The performance gap between the two stages is evident and gets even more emphatic for an increasing transmit SNR $p/N_{0}$. Further, it can be seen that PER of the $1^{\rm st}$ stage dominates the overall system performance for relatively low $\alpha_{1}$ values in comparison to PER at the $2^{\rm nd}$ stage; and vice versa for relatively high $\alpha_{1}$ values. This is a reasonable outcome since lower $\alpha_{1}$ values typically reduce $\gamma_{1}$ in \eqref{g1} as well as increase $\gamma_{2}$ in \eqref{g2}, while quite the opposite result holds as $\alpha_{1}$ increases.  

\begin{figure}[h]
\centering{\includegraphics[trim=0.0cm -0.0cm -0.0cm 0.0cm, clip=true,totalheight=0.45\textheight]{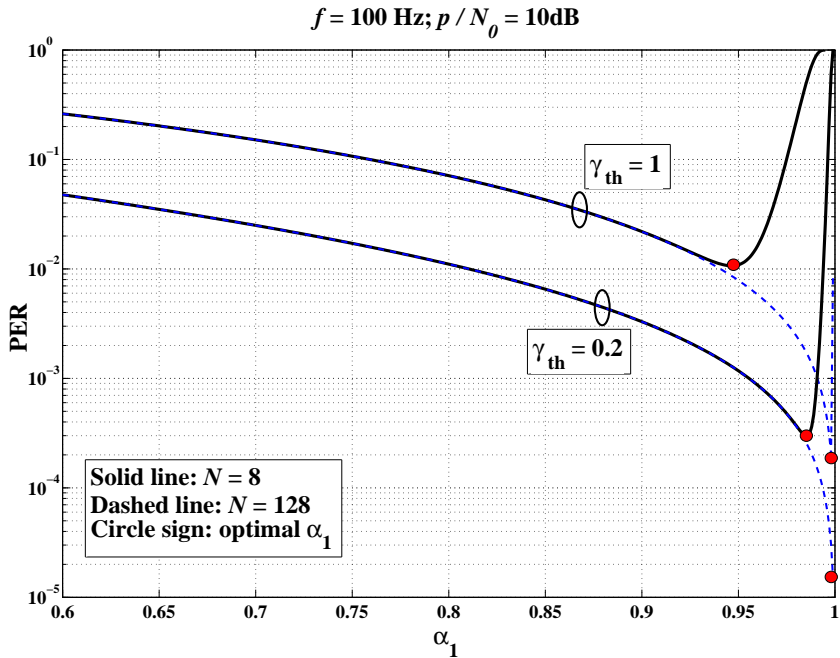}}
\caption{The union bound of the $2^{\rm nd}$ stage PER, i.e., $\overline{P^{(2)}_{s}}$, vs. various power allocation factors $\alpha_{1}$.}
\label{fig2}
\end{figure}

In Fig.~\ref{fig2}, the role of multiple antennas at the receiver is highlighted, by considering a conventional base station with $N=8$ antennas compared to a massive antenna array of $N=128$ antennas. The union bound of PER at the $2^{\rm nd}$ stage is illustrated. It is noteworthy that the massive antenna scenario outperforms the conventional case only when the power allocation factor $\alpha_{1}$ is carefully selected. Obviously, the PER performance is not affected by an increase of $N$ for a relatively low $\alpha_{1}$. On the other hand, there is a certain range of $\alpha_{1}$ values, where the presence of a vast antenna array is greatly beneficial. The optimal $\alpha_{1}$ (namely, $\alpha^{\star}$) is lying in this range, which can be quite easily computed as per $\mathcal{P}_{1}$. Also, as expected, PER is dramatically reduced for a lower rate threshold, $\gamma_{\rm th}$. Notice that a relatively low data rate requirement is the typical case of various URLLC applications since high data rates are being sacrificed so as to guarantee connectivity and ultra-high reliability.   

\section{Conclusion}
An uplink two-user NOMA communication system was considered, which operates under independent Rayleigh faded channels. The case when the users and receiver are equipped with a single-antenna and multi-antenna elements, respectively, was studied. Moreover, user mobility was supported reflecting several practical applications, e.g., vehicle-to-vehicle and device-to-device networking setups. Particularly, we focused on the short packet transmission regime, which relates to the rather timely URLLC applications. Closed-form expressions regarding the system PER were derived under arbitrary user mobility profiles and various antenna array ranges. Finally, some new useful engineering insights were obtained, while the optimal NOMA power allocation per user was formulated based on the derived PER results.

\end{document}